\documentclass[showpacs,prl,twocolumn,floatfix]{revtex4}
\usepackage{graphicx,float,amsmath}

\begin{document}

\title{Giant room-temperature piezoresistance in a metal/silicon hybrid}

\author{A.C.H. Rowe}
\email{alistair.rowe@polytechnique.edu}
\affiliation{Laboratoire de physique de la mati\`ere condens\'ee, \'Ecole Polytechnique, CNRS, 91128 Palaiseau Cedex, France}

\author{A. Donoso-Barrera}
\affiliation{London Centre for Nanotechnology and Centre for Mathematics and Physics in the 
Life Sciences and Experimental Biology (CoMPLEX), University College London, 17-19 Gordon Street, London WC1H 0AH, UK
}

\author{Ch. Renner}
\altaffiliation{Work partially carried out at the London Centre for Nanotechnology, UCL, 17-19 Gordon Street, London WC1H 0AH, UK
}
\affiliation{Department of Condensed Matter Physics and NCCR Materials with Novel Electronic Properties, University of Geneva, 24 Quai Ernest-Ansermet, CH-1211 Geneva 4, Switzerland}

\author{S. Arscott}
\affiliation{ Institut d'Electronique, de Micro\'electronique et de Nanotechnologie (IEMN), CNRS UMR8520, Avenue Poincar\'e, Cit\'e Scientifique, 59652 Villeneuve d'Ascq, France}

\begin{abstract} 
Metal/semiconductor hybrids are artificially created structures presenting novel properties not exhibited by either of the component materials alone. Here we present a giant piezoresistance effect in a hybrid formed from silicon and aluminum. The maximum piezoresistive gage factor (GF) of 843, measured at room temperature, compares with a GF of -93 measured in the bulk homogeneous silicon. This piezoresistance boost is not due to the silicon/aluminum interface, but results from a stress induced anisotropy in the silicon conductivity that acts to switch current away from the highly conductive aluminum for uniaxial tensile strains. Its magnitude is shown, via the calculation of hybrid resistivity weighting functions, to depend only on the geometrical arrangement of the component parts of the hybrid.

\end{abstract}

\maketitle

Interest in new materials presenting a large piezoresistive response is driven by applications involving the direct electrical sensing of mechanical stress, in particular in micro- and nano-electromechanical systems (MEMS/NEMS) \cite{elwenspoek01}. Examples from physics and biology include all-electrical atomic force microscopy \cite{tortonese93} and bioMEMS \cite{mckendry02} respectively. Giant piezoresistance at cryogenic temperatures or based on surface effects has been reported in a variety of materials \cite{aslam92, he06, tombler00, shkolnikov04, shekawat06}. A fundamentally different and unexplored route to higher piezoresistance is via the fabrication of hybrid structures combining metals and semiconductors similar to those which exhibit extraordinary magnetoresistance (EMR) in a perpendicular applied magnetic field \cite{solin00}. The qualitatively new properties exhibited by these metal/semiconductor hybrids (which are not observed in the component materials alone) are a result of artificial structuring rather than composition. In other words, the novel properties are a result of a new set of boundary conditions applied to the appropiate equation of motion, a situation encountered often in a wide variety of physical systems \cite{yablonovich91, levy90, mauroy04}. Here we present a giant piezoresistance effect in a metal/semiconductor hybrid composed of bulk silicon and aluminum that exhibits room temperature piezoresistive gage factors (GF) an order of magnitude larger than either of the constituent materials. It is shown that the piezoresistance boost is determined by the geometrical arrangement of the component parts.

Mechanical stress applied to a solid induces a change in the electrical resistance, $R = \frac{l}{\sigma A}$, via a change in the effective length ($l$) and cross sectional area ($A$). This change in dimension is the only source of piezoresistance in metals and results in a GF, defined as the relative resistance change per unit strain,  $\frac{\Delta R}{R\epsilon}$, typically of the order of 2. In semiconductors, particularly indirect band-gap semiconductors such as silicon, mechanical stress affects the electronic band structure thus modifying the effective electron mass, the mobility and $\sigma$ \cite{herring55}. The resulting change in resistance is significantly larger than the dimension change effect with GF $\approx$ 100 in silicon, dependent upon the doping, the direction of the mechanical stress, and the direction of current flow with respect to the crystal axes \cite{smith54, kanda82}. The magnitude of this physical piezoresistance, determined entirely by the piezoresistive coefficients ($\pi_{11}$, $\pi_{12}$ and $\pi_{44}$) in a cubic crystal such as silicon, is an immutable and intrinsic property of a homogeneous semiconductor. In the hybrids studied here the anisotropic physical piezoresistance of silicon under uni-axial tensile strain is used to deflect current away from a highly conductive, appropriately positioned aluminum shunt, thereby yielding a very large relative increase in resistance. The effect is shown to be qualitatively identical to the magnetic field induced deflection of current around a metallic inhomogeneity in InSb that gives rise to the EMR \cite{solin00}.

\begin{figure}
\includegraphics[clip,width=8 cm] {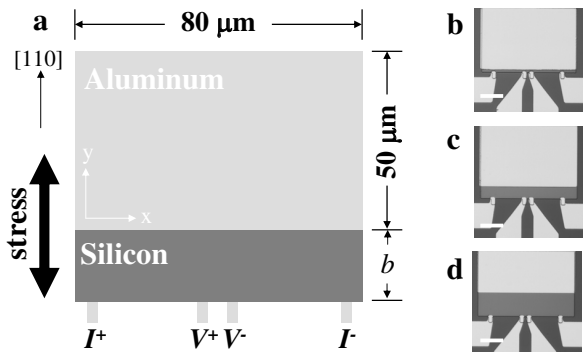}
\caption{(a) Schematic diagram of the aluminum/silicon hybrid. Stress is applied along the [110] crystal direction, perpendicular to the aluminum/silicon interface for a number of devices corresponding to different values of $b$. Photographs of three different geometries with (b) $b$ = 2 $\mu$m, (c) $b$ = 5 $\mu$m, and (d) $b$ = 20 $\mu$m. In each case the white scale bar is 20 $\mu$m long.}
\label{figure1}
\end{figure}

A schematic diagram of the hybrid, fabricated using standard lithography techniques from Boron implanted silicon (p = 1 $\times 10^{17}$ cm$^{-3}$, implant depth = 100 nm), is shown in Fig.\ref{figure1}a along with the crystal axes orientation. The dimension $b$ specifies the distance between the external leads for the four-terminal resistance measurement and the shunt and is equal to the length of the silicon part of the hybrid in the y-direction. A number of different geometries corresponding to $b$ = 1, 2 (Fig.\ref{figure1}b), 3, 4, 5 (Fig.\ref{figure1}c), 6, 7, 8, 9, 10, 12, 14, 16, 18 and 20 (Fig.\ref{figure1}d) microns were studied. In addition, shuntless control devices with $b$ = 20 $\mu$m were also fabricated. Standard four-terminal resistance measurements were made using the four 3 $\mu$m wide external leads: current is supplied to the outer leads and a voltage is measured across the inner leads. Individual 30 mm x 16 mm chips cut from the silicon wafer, each containing four geometrically different aluminum/silicon structures at its center on the chip surface, were mounted in a specifically designed apparatus for the application of uniaxial tensile strain along the [110] crystal axis using a bending technique similar to that described elsewhere \cite{rowe02}. Strain was measured using commercial metal foil strain gages glued to the rear face of the chip.

\begin{figure}
\includegraphics[clip,width=8 cm] {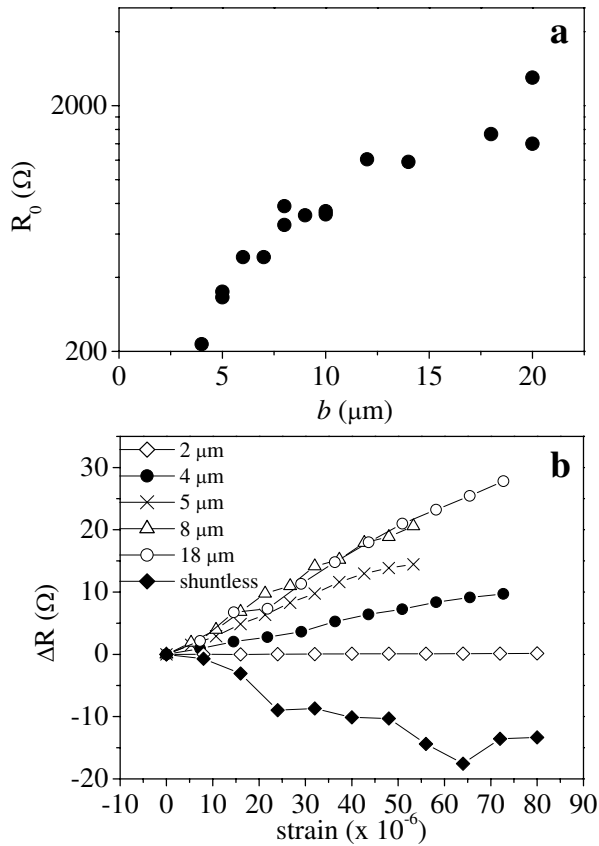}
\caption{(a) Dependence of the zero strain resistance on $b$. As $b$ is reduced, more current passes through the metal shunt thereby reducing the resistance. (b) Change in resistance,  $\Delta R$, under strain for a number of different device geometries. Depending on the device geometry $\Delta R$ can be large and positive (e.g. $b$ = 8 $\mu$m), close to zero ($b$ = 2 $\mu$m) or large and negative (shuntless control device).}
\label{figure2}
\end{figure}

\begin{figure}
\includegraphics[clip,width=8 cm] {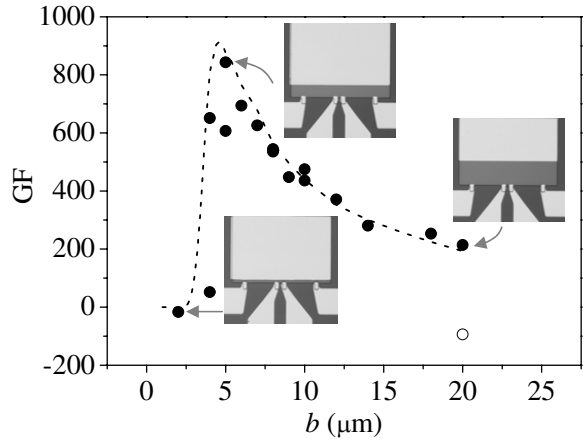}
\caption{Measured GF versus $b$, filled circles. A peak of 843 is observed at $b$ = 5 $\mu$m. This compares with a measured GF in the homogeneous silicon of -93 ($b$ = 20 $\mu$m shuntless control device), open circle. The solid horizontal line is the GF calculated for the homogenous silicon and the dotted curve is that calculated for the hybrid using only the non-adjustable parameters $\sigma_{0}$, $\sigma_{0,Al}$, $\pi_{11}$, $\pi_{12}$ and $\pi_{44}$.}
\label{figure3}
\end{figure}

As shown in Fig.\ref{figure2}a, the zero strain resistance increases monotonically as a function of $b$ since the metallic shunt more efficiently short circuits the silicon part of the device for small $b$ (the contact resistance was measured using a ladder network to be $1.6 \times 10^{-6}$ $\Omega$cm$^2$, consistent with the $10^{20}$ cm$^{-3}$ Boron doping used to form the external Ohmic contacts and the shunt contact). This is an important observation since blocking contacts (be they resistive or Schottky-like) will electrically isolate the aluminum from the silicon reducing the piezoresistance boost \cite{holz03b}. As for the shuntless control device, the piezoresistance in the metal/semiconductor hybrids is quasi-linear in strain (see Fig.\ref{figure2}b). The slope of each of these curves normalized to the zero strain resistance is the GF. A plot of the GF versus $b$ shows a peak of 843 at intermediate geometries ($b$ = 5 $\mu$m) (see Fig.\ref{figure3}) which resembles the peak observed in the EMR at intermediate geometries \cite{solin00}. For $b < 5$ $\mu$m the GF drops off very rapidly to zero (within the error bars of the measurement). At these geometries the arrangement of current and voltage leads ensures that it is mainly the metal which contributes to the resistance, so it is the intrinsic piezoresistance of the latter (which for these strains is negligibly small) that determines the measured GF. At large values of $b$, the GF drops off more slowly as the piezoresistive properties of the bulk semiconductor begin to dominate. Interestingly, the shuntless device (with $b$ = 20 $\mu$m) has a negative GF as is to be expected for p-type silicon in this configuration \cite{kanda82}. The presence of the metallic shunt thus results in both a change in magnitude and sign of the GF.  

To understand the piezoresistive behaviour, the piezoconductivity tensor, $\tensor{\sigma}$, for silicon must be considered \cite{smith54}. In a cubic crystal with applied stress, $X$, along the [110] direction, the components of this tensor are $\sigma_{xx} = \sigma_0 ( 1 - \frac{X}{2} [\pi_{11} + \pi_{12} - \pi_{44} ])$ and $\sigma_{yy} = \sigma_0 ( 1 - \frac{X}{2} [\pi_{11} + \pi_{12} + \pi_{44} ])$ with $\sigma_{xy} = \sigma_{yx} = 0$. Here, $\sigma_0$ is the isotropic zero stress silicon conductivity. In p-type silicon for the moderate doping density considered here  $\pi_{11} = 6.6 \times 10^{-11}$ Pa$^{-1}$,  $\pi_{12} = -1.1 \times 10^{-11}$ Pa$^{-1}$ and $\pi_{44} = 138.1 \times 10^{-11}$ Pa$^{-1}$ respectively \cite{smith54}. Note that the stress, $X = E_{110}$ where $E_{110} \approx$ 170 GPa is Young's modulus in silicon, is taken positive for tensile strains. In the notation used here the x-direction corresponds to the [1$\bar{1}$0] crystal direction and the y-direction to the [110] crystal direction parallel to $X$. The current density is then written as $\vec{J} =  \tensor{\sigma} \vec{F}$ where $\vec{F}$ is the electric field vector. At zero stress ($X$ = 0) the piezoconductivity tensor is diagonal and the conduction is described by the Drude model. At non-zero strain the silicon conductivity becomes anisotropic and, for the tensile strains considered here, increases in the x-direction and decreases in the y-direction. In other words the current has more difficulty reaching the metallic shunt under strain while it passes more easily along the semiconducting portion of the hybrid. In effect the strain partially electrically isolates the shunt, resulting in a very large positive GF since its conductivity is orders of magnitude greater than that of the semiconductor. On the other hand, because of its elongated form, the resistance of the shuntless device is dominated by the silicon conductivity in the x-direction with a corresponding negative GF. The GF versus $b$ curve is calculated by solving Laplace's equation in two dimensions using finite element techniques with the measured value of $\sigma_{0}$ = 26 ($\Omega$cm$)^{-1}$, and an aluminum conductivity $\sigma_{0,Al} = 3.8 \times 10^5 (\Omega$cm$)^{-1}$ along with the piezoresistance coefficients cited above. No account is taken of stress induced dimension changes since these are negligible. The resulting curve is shown as a dotted line in Fig.\ref{figure3}. The remarkable agreement between the data and the model (which contains no adjustable parameters) indicates that the piezoresistance boost is in no way related to stress induced changes at the contact region between the silicon and the aluminum \cite{rowe03}, but is a true bulk effect.

It is of interest to compare and contrast the piezoresistance boost to the EMR also observed in planar metal/semiconductor hybrids \cite{solin00, zhou01b}. In EMR, a perpendicular magnetic field deflects current away from the metallic shunt via the Lorenz force, thereby greatly increasing the resistance. In a qualitative sense, the piezoresistance boost described here is the same. Accounting for its crystal orientation dependence, tensile strain deflects current away from the metallic shunt thereby increasing resistance. However, unlike EMR which is symmetric in magnetic field, the piezoresistance effect is anti-symmetric in strain, meaning that for the present geometry, current can be deflected into the shunt for compressive strains. These differences are quantitatively expressed by the symmetry of the magneto- and piezo-conductivity tensors. The former corresponds to a global rotation of the current with respect to the electric field by the Hall angle, whereas the latter corresponds to a rotation whose angle depends on the local direction of the electric field. 

\begin{figure}
\includegraphics[clip,width=8 cm] {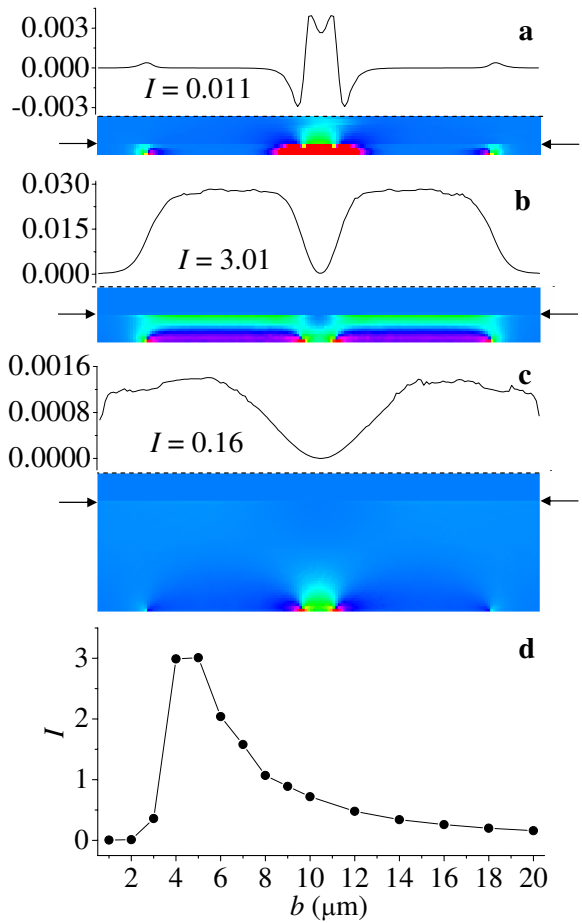}
\caption{Finite element calculation of the HRWF for three geometries, (a) $b$ = 2 $\mu$m, (b) 5 $\mu$m and (c) 20 $\mu$m. To better view the interface, only the first five microns of the metal above the interface (indicated by the black arrows) are shown. In each case, red (light blue) areas contribute strongly (weakly) to the resistance. For $b$ = 2 $\mu$m, the metal strongly contributes to the resistance since it is close to the external contacts. For $b$ = 5 $\mu$m and $b$ = 20 $\mu$m the metal no longer contributes and in the latter case, the semiconductor dominates the resistance. The interfacial layer contributes the most to the resistance at $b$ = 5 $\mu$m (green shading) as confirmed by the integral, $I$, of the HRWF line scan at the interface (d). At this value of $b$, a stress induced redirection of the current in the silicon will most effectively switch current from the metal to the semiconductor resulting in the largest possible piezoresistance.}
\label{figure4}
\end{figure}

The hybrid resistivity weighting function (HRWF) provides a quantitative measure of the contribution that each point in the hybrid makes to the zero stress resistance \cite{rowe05}; its value at any given point in space depends on the device geometry together with the current and voltage lead positions. Fig.\ref{figure4}a-c shows a sample of the HRWFs for three of the geometries tested here ($b$ = 2 $\mu$m, 5 $\mu$m and 20 $\mu$m) along with the line scans of the HRWF and its integral, $I$, at the interfacial region. The HRWF is calculated using a smoothed, flat-topped Heaviside function 0.5 $\mu$m $\times$ 0.5 $\mu$m in size and $5 \times 10^4$ $(\Omega$cm$)^{-1}$ in amplitude as the conductivity perturbation. In the color plots the red (light blue) regions contribute strongly (weakly) to the measured resistance. The presence of green (representing a mild positive contribution to the resistance) clearly indicates that the interface region is most important for $b$ = 5 $\mu$m, and this is quantitatively established in the plot of $I$ versus $b$ in Fig.\ref{figure4}d. The peak in $I$ at the optimal geometry for the piezoresistance boost is no coincidence: a strain-induced deflection of the current will most effectively switch current to/from the aluminum (and thus change the resistance) when the current grazes the silicon/aluminum interface at zero strain. The geometry is thus predominant in determining the novel properties (i.e. the giant piezoresistance) of the hybrid which are not observed in the component materials alone.

While the giant piezoresistance obtained by artificially structuring metal/semiconductor hybrids represents a fundamentally new route towards ultra-sensitive, silicon compatible strain gages, the effect should not be limited to combinations of silicon and aluminum. Any combination of conducting materials, at least one of which presents an anisotropic piezoconductivity tensor, will do. Furthermore, there is a vast geometric parameter space over which the piezoresistive properties can be tuned and it is unlikely that the simple rectangular geometries studied here will prove optimal for maximum piezoresistive response. 

\section{Acknowledgements}

This work was supported by the Agence National de la Recherche (ANR), project MEMSH ANR-06-NANO-037-01. ADB and CR acknowledge financial support from CoMPLEX.

\bibliographystyle{apsrev}
\bibliography{bibrowe}

\end{document}